\documentclass[traditabstract]{aa} 
\usepackage{graphicx}
\usepackage{txfonts}
\usepackage{natbib}
\bibpunct{(}{)}{;}{a}{}{,} 
\begin{document}
   \title{The Solar Oxygen Abundance from an Empirical Three-Dimensional Model}

   \author{H. Socas-Navarro\inst{1, 2}
   }

   \institute{Instituto de Astrof\'\i sica de Canarias,
     Avda V\'\i a L\'actea S/N, La Laguna E-38200, Tenerife, Spain
   \and
   Departamento de Astrof\'\i sica, Universidad de La Laguna, 38205, 
   La Laguna, Tenerife, Spain 
 }

   \date{}

   \authorrunning{Socas-Navarro}
   \titlerunning{Solar Oxygen Abundance}

\newcommand {\FeI} {\ion{Fe}{i}}
\newcommand {\OI} {\ion{O}{i}}
\newcommand {\ScII} {\ion{Sc}{ii}}
\newcommand {\NiI} {\ion{Ni}{i}}
\newcommand {\ltau} {$\log(\tau_{500})$}
\newcommand {\loge} {$\log \epsilon$}
\newcommand {\leO} {$\log \epsilon$(O)}
\newcommand {\leNi} {$\log \epsilon$(Ni)}
\newcommand {\leFe} {$\log \epsilon$(Fe)}
\newcommand {\leSc} {$\log \epsilon$(Sc)}

\abstract{The Oxygen abundance in the solar photosphere, and
  consequently the solar metallicity itself, is still a controversial
  issue with far-reaching implications in many areas of
  Astrophysics. This paper presents a new determination obtained by
  fitting the forbidden \OI \ line at 6300 \AA \ with an {\em
    observational} 3D model. The approach presented here is novel
  because previous determinations were based either on 1D empirical
  stratifications or on 3D theoretical models. The resulting best-fit
  abundances are \leO=8.90 and \leNi=6.15. Nevertheless, introducing
  minor tweaks in the model and the procedure, it is possible to
  retrieve very different values, even down to \leO=8.70.  This
  extreme sensitivity of the abundance to possible systematic effects
  is not something specific to this particular work but probably
  reflects the real uncertainty inherent to all abundance
  determinations based on a prescribed model atmosphere.  }


   \maketitle
%

\section{Introduction}
\label{sec:intro}

Oxygen is, after Hydrogen and Helium, the third most abundant element
in the Universe. A precise knowledge of its abundance in stellar
interiors is of critical importance in many areas of astrophysics
because of its contribution to opacity, free electrons and also
because it serves as a reference for other relevant elements that
cannot be directly measured in the photospheric
spectrum. Unfortunately, there are reasons to believe that our
knowledge of even the solar oxygen abundance is still far from being
precise. Several factors conspire to make it so elusive. First of all,
Oxygen is a very volatile element and therefore its meteoritic abundance
is not a valid proxy. Second, there are very few atomic 
indicators in the photospheric solar spectrum and they are all either
extremely weak forbidden lines or complex transitions suffering of
non-LTE effects and affected of additional uncertainties in the relevant
atomic processes.

In spite of the uncertainties, the prevailing paradigm that emerged
since the 1970s was that the solar Oxygen abundace was \leO\ in the
range between~8.8 and~8.9, in the logarithmic scale commonly used in
Astrophysics where H has a reference value of 12 (e.g., \citealt{C73,
  AG89, GS98}). In all cases, the procedure to determine chemical
abundances was always the same: From a prescribed model atmosphere,
one computes synthetic spectral lines and the abundances are adjusted
until a satisfactory agreement is found with some atlas
observations. The models employed were one-dimensional (1D), usually
derived empirically by fitting a large number of lines and continua,
and the photospheric observations were unresolved in space and time.

The paradigm was called under question by the controversial paper of
\cite{AGS+04} (hereafter AGSAK), claiming that the solar O abundance
needed to be revised downwards by almost a factor of two (\leO=8.66). The
procedure was conceptually very similar to what had been done before
except that, besides updated atomic parameters, the authors employed a
three-dimensional (3D) model resulting from a hydrodynamical numerical
simulation. Obviously, such large revision would have a considerable
impact on many areas of Astrophysics. In particular, it would ruin the
excellent agreement existing between solar interior models and
measurements from Helioseismology. Attempts to reconcile the models
with the new abundances have been unsuccessful thus far (e.g.,
\citealt{BA08} and references therein; \citealt{BA13}).

The debate on whether the proposed revision should be adopted has been
very intense and several papers have been published with new abundance
determinations, including those of \cite{APK06, SNN07, A08, CSN08,
  CLS+08, SAG+09, PAK09, CLS+11}. The frequency of publications on the
so-called {\em solar oxygen crisis} seems to have declined in recent
years, not because the issue has been satisfactorily resolved but
probably because there are no new arguments or data in favor of one
view or the other. Whatever the final outcome is, the work of AGSAK
has been of great importance because, at the very least, it has blown
the whistle on a grossly overlooked problem, namely how much can we
trust our abundance determinations. The choice of a suitable
atmospheric model is critical and it is not yet clear what the model
uncertainties are or how those uncertainties propagate into the final
result. \cite{A08} correctly remarks that, while a 3D atmosphere is
preferrable over a 1D, it is not clear that a model resulting from a
numerical simulation is better than an empirical one when it is
employed to fit observations. In addition, the work of \cite{CLS+08,
  CLS+11} demonstrates that going to 3D does not necessarily result in
lower abundances.

The present paper is an attempt to resolve the dilemma of theoretical
3D versus empirical 1D by taking the best of both approaches. A 3D
model obtained from observations is used to derive the O and Ni
abundances in the conventional way (i.e., fitting unresolved atlas
observations). As in my previous works on this subject
(\citealt{SNN07, CSN08}), the focus is not so much the final result
but the introduction of a novel methodology, hoping that it might open
a new path for better, more robust determinations in the future.

\section{Observations and the model}
\label{sec:model}

The 3D model was derived from a reanalysis of the data used in
\citeauthor{SN11} (\citeyear{SN11}, hereafter SN11) with a slightly
different approach and a new version of the code NICOLE
\citep{SNdlCRAA+14}. The new version incorporates several improvements
in accuracy and stability that make the model smoother and with a
lower {\em inversion noise} (the pixel-to-pixel fluctuation exhibited
by the retrieved parameters, such as the photospheric
temperature). Some other improvements include:

\begin{itemize}
\item The new NICOLE version supports 2-component inversions. Pixels
  that exhibit polarization signal above a certain threshold (32\% of
  the total) are inverted here with a full 2-component treatment,
  which is more consistent than the approach taken in the SN11
  paper.
\item A regularization term is added to the $\chi^2$ merit function to
  favor (when possible) smoother solutions over those with spurious
  high-frequency vertical fluctuations.
\item A Bezier spline interpolation scheme is now used to reconstruct
  the model atmosphere from the node values. Previously, the model was
  constructed from linear segments between the nodes. The new scheme
  results in smoother height runs of the retrieved physical
  parameters, thus getting rid of unsightly sharp corners and at the
  same time minimizing the possibility of overshooting between the
  nodes. For stability reasons, it is important to keep the model
  within reasonable ranges of the various physical parameters at every
  step of the iterative procedure. The algorithm sets constrains on
  the node values but the interpolation between nodes might overshoot
  beyond the safe range. This risk is minimized by the use of Bezier
  interpolation, which behaves similarly to splines with tension.
\item Includes hyperfine structure arising from the angular momentum
  coupling between electron and nuclear spin.
\end{itemize}

The dataset from which the model has been derived is exactly the same
as in SN11. A brief description is provided here for a sake of
completeness but the reader is referred to that paper for more
details. The data come from the Spectro-Polarimeter (SP) of the Hinode
satellite's Solar Optical Telescope (SOT;
\citealt{Hinode1,Hinode2,Hinode3,Hinode4,Hinode5}), particularly from
a quiet Sun observation acquired at UT 19:32:10 on 2007 September 24.

The SP scan spans the wavelength range between 6300.89 and 6303.29 \AA
\ with a sampling of 21.4 m\AA . Unfortunately the O/Ni blend at
6300.27 \AA \ that is the target of this study is just outside the
observed range. It is therefore impossible to perform here a
pixel-by-pixel abundance analysis {\it a la} \cite{SNN07}, which would
be of great interest e.g., to gauge with high confidence and precision
the total error of the entire procedure. Instead, element abundances
are derived here by fitting the disk-center average intensity from the
Kitt Peak Fourier Transfor Spectrometer (FTS) atlas of \cite{NL84},
following the traditional procedure of synthesizing line profiles from
the model and taking the spatial average of the synthetic spectra to
compare with the atlas.  Absolute wavelengths were obtained by
comparing the average observed spectrum to the FTS atlas. The spectral
point spread function of the Hinode SP is known and has been
considered in the analysis by applying it on the emergent synthetic
spectra. This means that, at each iterative step, the synthetic
profiles produced by the proposed model are convolved with the Hinode
PSF before being compared to the observed ones.

The field of view observed by the Hinode SOT is a very quiet region at
disk center. Spatial sampling is about 0.15'', which is approximately
half of the actual spatial resolution given by the SOT point spread
function. Standard flatfielding and polarimetric calibration
procedures were applied to the spectra. Some other corrections were
applied {\em a posteriori} as detailed in SN11.

The inversion fits both \FeI \ lines at 6301.5 \AA \ and 6302.5 \AA
\ simultaneously. There are some small departure from LTE
\citep{STB01} which are taken into account in the procedure using the
approximation described in SN11. Three different inversions were
carried out varying the assumed Fe abundance (7.40, 7.45 and 7.50,
respectively), resulting in three slightly different 3D models. This
will be helpful to gauge the influence of our uncertainties in the
model on the abundances obtained. The atomic line data employed are
listed in Table~\ref{table:atomic}. They are mostly those resulting
from a query to the VALD database (\citealt{PKR+95};
\citealt{KPR+99}; \citealt{KRP+00}), with the following remarks. The
oscillator strength for \FeI \ 6301.5 is from \cite{BKK91}. In the
absence of actual laboratory data for the 6302.5 line, the $\log(gf)$
derived empirically by SN11 is used here. For both \FeI \ lines,
collisional broadening is treated using the method of
\cite{AO95}. The broadening constants, listed in
Table~\ref{table:atomic}, are obtained with the code of
\cite{BAOM98}. The wavelength correction (20 m\AA \ to the blue)
applied to the \ScII \ line by \citet{A08} is necessary here, as
well. For the $\log(gf)$, however, no further adjustment was
needed. The original line strength from the VALD database combined
with the meteoritic Sc abundance yields a satisfactory fit to the
line shape, as shown below. Oscillator strengths for the two relevant
Ni isotopes (treated here as two blended lines with a slight
wavelength shift) are from \cite{JLL+03}. Finally, for the forbidded
\OI \ line, the value adopted is that from \cite{SZ00}, as recommended
by \cite{SAG+09}.

As noted above, the inversion is done differently depending on whether
the spatial pixel to invert is magnetic or not. Magnetic pixels are
defined as those that exhibit polarization signal above the $3-\sigma$
noise level in any one of the Stokes profiles $Q$, $U$ or $V$. In both
cases the model is initialized with the HSRA as starting guess
\citep{GNK+71} and then iterated in two successive cycles. The number
of nodes used in the inversions for the various physical parameters is
listed in Table~\ref{table:nodes}. Note that the inversion
corresponding to the second cycle of magnetic pixels uses a
2-component model and it has a different number of nodes for each
component. In those cases (magnetic pixels), the atmosphere is
parameterized as consisting of a magnetic component (Comp 2 in the
table) coexisting with some filling factor inside the pixel with a
non-magnetic surrounding (Comp 1).

\begin{figure}
  \centering
  \includegraphics[width=0.5\textwidth]{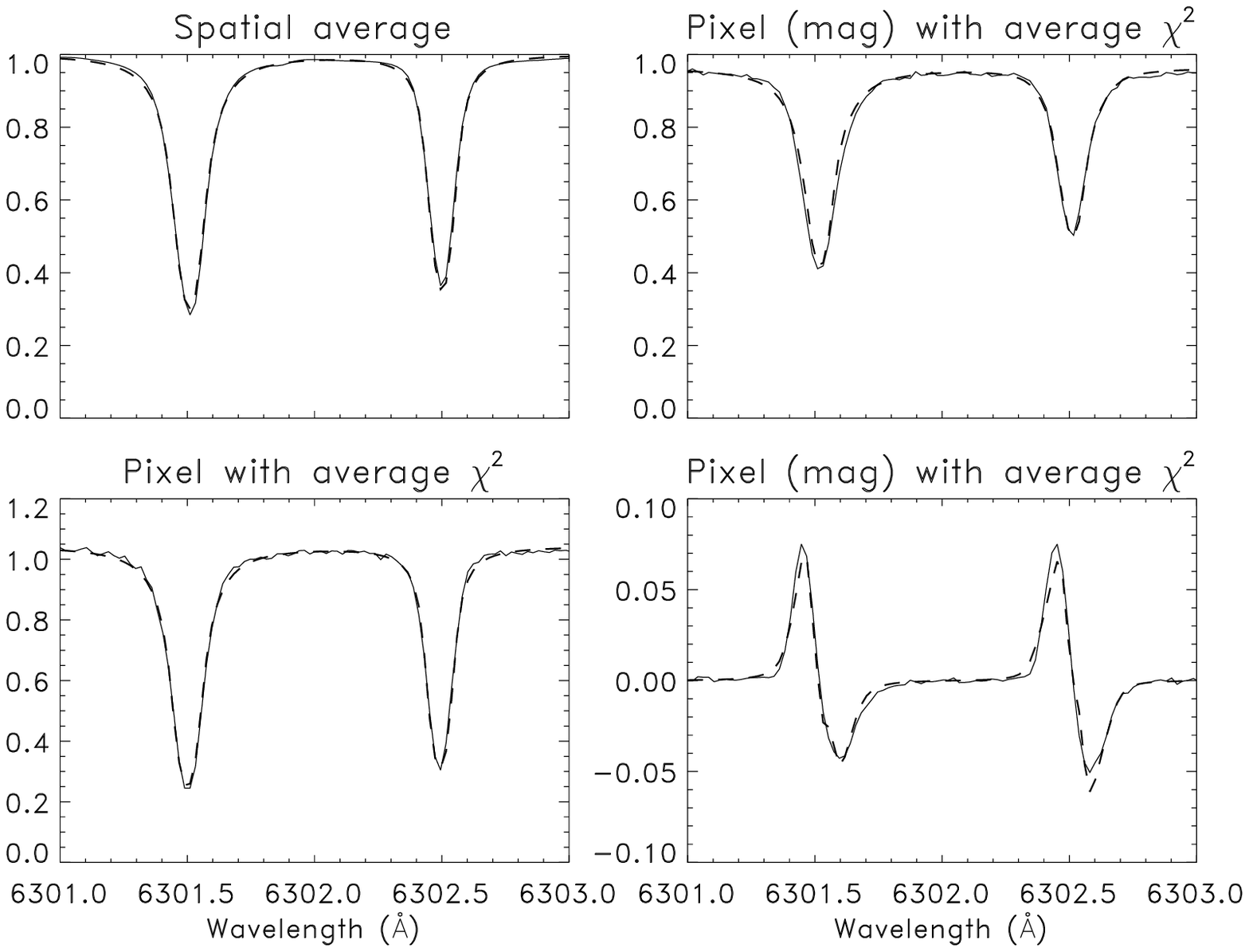}
  \caption{Some representative fits obtained with the empirical 3D
    model. In all panels the solid line is the Hinode observations and
    the dashed line is the synthetic profile computed with
    NICOLE. Upper left: Spatial average over the entire field of
    view. Lower left: Fit to a non-magnetic profile whose $\chi^2$ is
    similar to the average $\chi^2$ in all the non-magnetic
    pixels. The quality of this fit may then be considered as typical
    of the entire region. Right panels: Fits to Stokes~$I$ (upper)
    and~$V$ (lower) profiles from a pixel with magnetic signal whose
    $\chi^2$ is similar to the average $\chi^2$ in all the magnetic
    pixels. The quality of this fit may then be considered as typical
    of the entire region.}
  \label{fig:fits}%
\end{figure}

\begin{figure*}
  \centering
  \includegraphics[width=0.8\textwidth,height=\textheight]{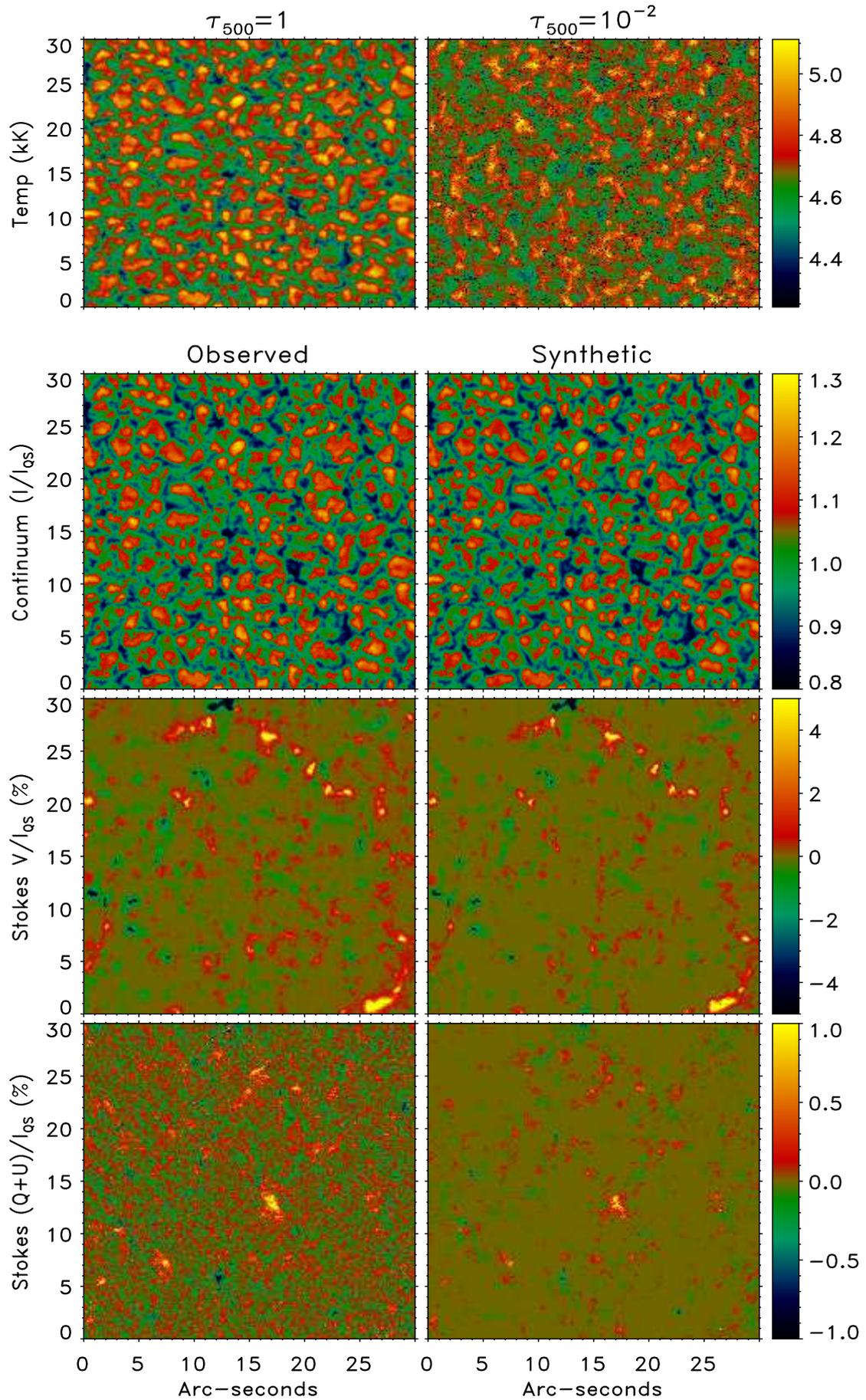}
  \caption{Spatial distribution of various magnitudes derived from the
    empirical 3D model used in this work. Top panels: Horizontal
    cuts of the temperature at the base of the photosphere (left) and
    at the height where \ltau$=10^{-2}$ (right). The rest of the
    panels show a comparison between various observed (left) and
    synthetic (right) quantities.}
  \label{fig:maps}%
\end{figure*}

Some representative fits are shown in Figure~\ref{fig:fits} to give an
idea of how accurately the synthetic lines match the
observations. Perhaps the most relevant plot for the discussion here
is the top-left panel, which shows the spatially-averaged
profiles. Notice that this is not actually a fit but the result of
combining all the individual synthetic profiles together and comparing
it to the average of the observations. The other three panels show
typical fits in magnetic and non-magnetic pixels.

Figure~\ref{fig:maps} illustrates the spatial distribution of some
sample quantities. No effort has been made to put the model on a
common geometrical scale since that is irrelevant for the purpose of
computing the emerging profile from each model column. Therefore, the
reference height scale employed in this paper is the monochromatic
continuum optical depth at 500~nm, \ltau.

As with the earlier version, the new 3D model is publicly available
and may be downloaded from:
\begin{verbatim}
ftp://download:data@ftp.iac.es/
\end{verbatim} 
The files are licensed under the GPLv3 general public
license\footnote{See http://www.gnu.org/licenses/gpl.html} which
explicitly grants permission to copy, modify (with proper credit to
the original source and explanation of the modifications) and
redistribute it.

\section{Analysis}
\label{sec:analysis}

\subsection{The \ScII \ line and the missing dynamics}
\label{sec:scii}

Very close to the O/Ni blend that constitutes the main target of this
study lies another interesting spectral feature of similarly weak
nature, emerging from a \ScII \ transition. This line is of great
interest for reasons that shall be explained below and has been
included in the synthetic spectrum computed from the 3D model. In
order to calculate the \ScII \ profile one needs to consider its
hyperfine structure. Fortunately, accurate atomic parameters exist in
the literature (e.g., in the VALD database) and one can easily feed
those values into NICOLE to produce realistic line shapes.

The inclusion of hyperfine structure for the \ScII \ line slows down
the calculation by nearly a factor 100. As a result, computing the
200$\times$200 columns of the 3D model takes a significant (but still
feasible) amount of time. However, it quickly becomes prohibitive to
repeat the full model synthesis for each point in the three-parameter
grid of abundances of Fe, Ni and O required for the abundance
determination in Sect~\ref{sec:abundances} below. It is then
convenient, if not entirely necessary, to find computationally cheaper
alternatives.

\begin{figure}
  \centering
  \includegraphics[width=0.5\textwidth]{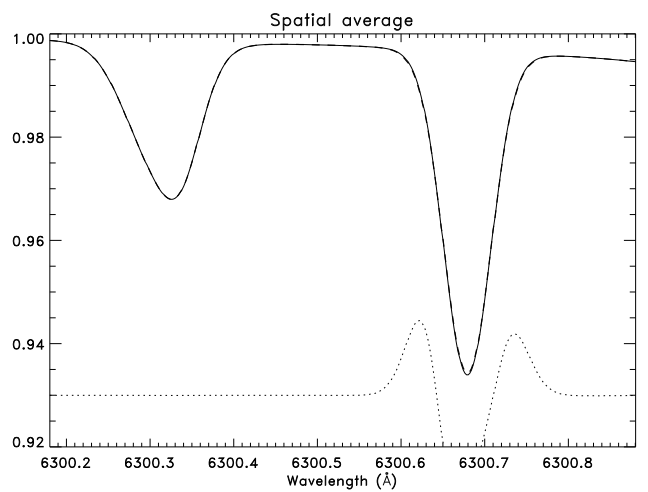}
  \caption{Comparison of hyperfine splitting effects (solid) and an
    {\em ad hoc} broadening (dashed) of 0.8~km~s$^{-1}$ on the average
    profile of the \ScII \ spectral line. The dotted line shows the
    difference between both profiles, multiplied by a factor 50 and
    conveniently offset for the sake of easy visualization.}
  \label{fig:hf_vmic}%
\end{figure}

The overall effect of the hyperfine structure on the {\em spatially
  averaged} \ScII \ intensity profile is simply a broadening of the
line. With some experimentation, it is straightforward to see that the
average \ScII \ synthetic profile emerging from the 3D model is nearly
identical to one computed without hyperfine structure but adding a
microturbulent velocity of 0.8~km~s$^{-1}$, as seen in
Fig~(\ref{fig:hf_vmic}). The extra parameters, in addition to those in
Table~\ref{table:atomic}, needed for the hyperfine structure calculation
  (\citealt{YCD+88}; \citealt{AHN+82}) are: I=3.5,
A$_{low}$=-27.9, B$_{low}$=19,
A$_{up}$=125.4, B$_{up}$=7, with all A and B constants in MHz. In
view of this result, all the calculations shown in the remainder of
this paper have been carried out simulating the effects of hyperfine
structure in the \ScII \ line by applying an ``artificial''
microturbulence of 0.8~km~s$^{-1}$ (only to this line!), which is a
very good approximation for the spatially averaged intensity.

\begin{figure}
  \centering
  \includegraphics[width=0.5\textwidth]{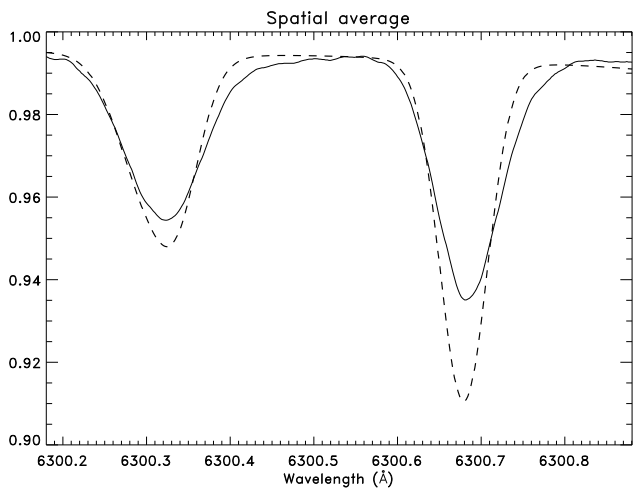}
  \caption{The synthetic \ScII \ line at 6300.68 \AA \ (dashed) is too
    narrow when compared to the observed atlas (solid) for any
    combination of abundance and oscillator strength.}
  \label{fig:firstfit}%
\end{figure}

A first comparison of the synthetic spectrum to the atlas observation,
plotted in Fig~(\ref{fig:firstfit}), shows that it is possible to
reach a reasonable agreement in the O/Ni blend but a very poor fit to
the \ScII \ line. The abundance values for the calculation in the
figure are \leFe=7.50, \leO=8.80, \leNi=6.20. The discrepancy of the
\ScII \ feature is too large to be reconciled by tweaking its
abundance or the atomic parameters. It is possible to make the line
core deeper by increasing its abundance or oscillator strength but it
would still be too narrow compared to the observed profile.

In order to bring the synthetic \ScII \ line into agreement with the
observation, it is necessary to add some additional broadening. The
natural explanation for the missing broadening is to assume that, even
with Hinode's high spatial resolution, there are still some
small-scale plasma flows at the base of the photosphere that are not
being sufficiently resolved in the observations. It is important to
note that the \FeI \ lines are very well reproduced without requiring
any microturbulence, which indicates that most of the small-scale
dynamics seen by these lines has been properly captured in the 3D
model. Therefore, the missing dynamics is probably located in the
deeper layers, where the O/Ni and the \ScII \ lines form but the \FeI
\ lines have very little sensitivity.

A comparison of the 3D model employed here with the hydrodynamical
simulation of AGSAK further supports the notion of the missing
dynamics scenario. The {\em rms} spread of velocities at the \ltau$=1$
level is much larger in the simulation than in the 3D empirical
model. With the information provided by the \ScII \ line, it is
straightforward to correct this problem by applying an {\em ad-hoc}
enhancement factor v$_{enh}$ to the velocities in the model and
tweaking this factor until the line is properly fitted. Since the
\ScII \ line has a very similar strength and formation region to the
O/Ni blend, this enhancement will also correct any possible missing
broadening in the O and Ni features. After some experimentation, it
was found that the optimal {\em rms} spread for the \ScII \ line is
v$_{rms}$=1.36~km~s$^{-1}$, about a factor of 2 greater than in the
empirical model. This figure is considerably lower than the
2.47~km~s$^{-1}$ of the AGSAK model. In fact, adopting their value of
the {\em rms} velocity spread would result in {\em excessive}
broadening of the observed \ScII \ line, incompatible with the
observations. With the correction described here, the agreement
between observed and average synthetic profiles improves enormously, as
shown below.

\section{Abundance determinations}
\label{sec:abundances}

Having a 3D model that accurately reproduces the \ScII \ line, we can
now proceed with confidence to the derivation of the O and Ni
abundance by the usual procedure. One computes the synthetic profiles
of the O/Ni blend for an array of [\leO, \leNi] values at each model
column, takes the spatial average over the entire field of view and
compares the result to the observation in order to find which pair of
abundances produces the best match. The comparison is performed by
defining, for each set of abundances, a $\chi^2$ function as the sum
of the quadratic difference between synthetic and observed profiles
for all wavelengths in the range. 

Since there are actually three slightly different 3D models, resulting
from using three different Fe abundances (\leFe=7.40, 7.45 and 7.50)
in the inversion of the Hinode \FeI \ profiles, we perform the
comparison for all three cases, resulting in an effective
three-dimensional grid of abundances (\leFe, \leO, \leNi). Even though
the best value for the solar Fe abundance is probably \leFe=7.50, as
concluded by previous works \citep{STB01}, it is nevertheless
interesting to use the three inversion results as this will hopefully
provide some insight into how slight changes in the model can influence
the inferred abundances.

\begin{figure}
  \centering
  \includegraphics[width=0.5\textwidth]{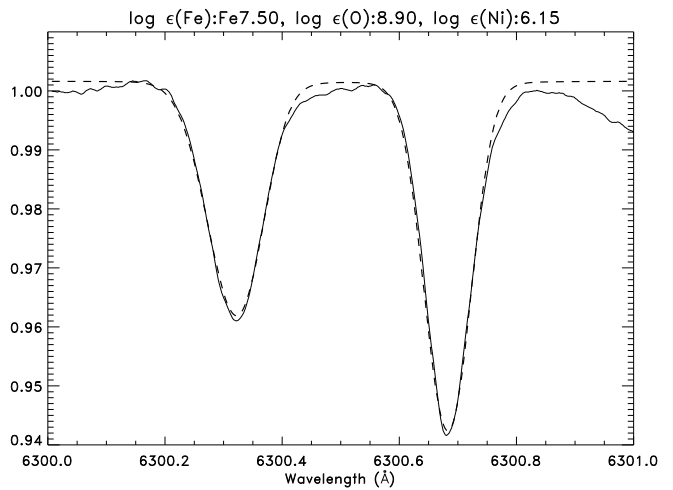}
  \caption{Best fit to the [O/Ni] feature in the abundance grid.}
  \label{fig:bestfit}%
\end{figure}

The profiles plotted in Fig~\ref{fig:bestfit} represent the best
fit. The corresponding triplet of abundance values is \leFe=7.50,
\leO=8.90 and \leNi=6.15. As a side note, the Sc abundance required to
fit this line is the meteoritic value of 3.05.

\begin{figure}
  \centering
  \includegraphics[width=0.5\textwidth]{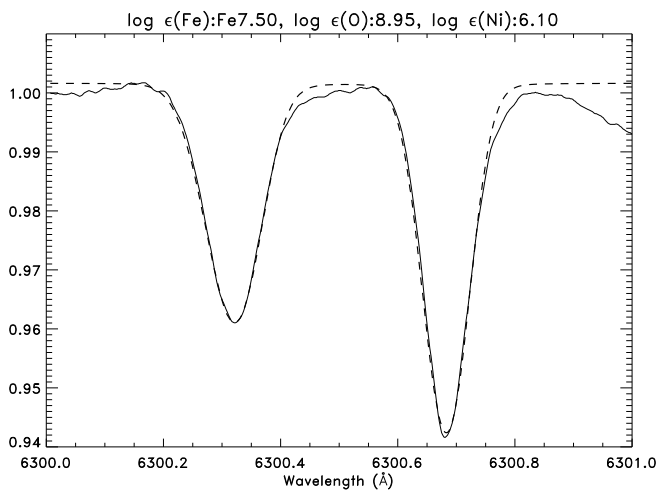}
  \caption{Second best fit to the [O/Ni] feature in the abundance grid.}
  \label{fig:secondbestfit}%
\end{figure}

To illustrate how the fit degrades as one moves away from the optimal
set of abundances, Fig~\ref{fig:secondbestfit} shows the second best
fit in the grid. This new fit was obtained with \leO=8.95 and
\leNi=6.10 and the resulting $\chi^2$ is only 20\% higher than
before. We can see in the figure that that fit is remarkably tight,
even though the difference in O abundance is rather significant. A
comparison of the results obatined with all three models
(corresponding to \leFe=7.40, 7.45 and 7.50) shows that the best fit
is always achieved with \leO=8.90 \ and \leNi \ varying between 6.10
and 6.15, in all three cases with similarly good fits to those shown
in Fig~\ref{fig:bestfit}. If we also include the second best match in
the comparison, we have three more values to consider. As before, the
$\chi^2$ is only slightly worse than for the best (30\% higher, at the
most) and the fits look visually very similar to those in
Fig~\ref{fig:bestfit}. In this case, the spread of good fit abundances
increases to the range from 8.85 to 8.95 for \leO \ and remains
between 6.10 and 6.15 for \leNi .

In order to further explore the sensitivity of the abundance results
to the model employed for the calculation, the same procedure was
applied to a slightly perturbed version of the \leFe=7.50 model. The
perturbation consists of a linear addition to the temperature in the
range of optical depths \ltau=(-1,1) and has a total amplitude of
50~K, going from +25~K at \ltau=1 down to -25~K at \ltau=-1. With such
small perturbation, the inferred abundances of O and Ni change to 8.95
and 6.10, respectively.

\begin{figure}
  \centering
  \includegraphics[width=0.5\textwidth]{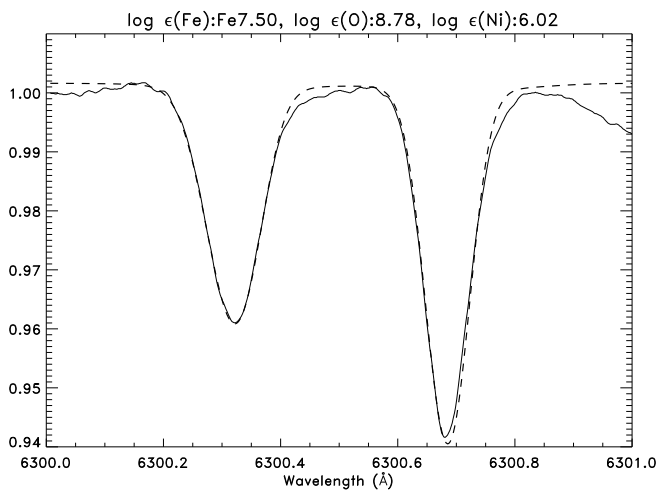}
  \caption{Best fit abundances using a slightly perturbed model with a
    perturbation in temperature ranging from 50~K at \ltau=1 to -50~K
    at \ltau=-1.}
  \label{fig:perturbedbest}%
\end{figure}

Other factors that affect the result are the continuum reference and
the wavelength calibration. If the continuum is chosen slightly lower
or higher, both O and Ni abundances will change accordingly. This is a
common problem in all abundance determinations. Similarly, a slight
wavelength shift of a few m\AA \, might be compensated by altering the
ratio of O and Ni in the blend to recover a good fit. Finally, the
line broadening produced by the dynamics going on in the lower
photosphere has a significant impact, as well. The AGSAK dynamics is
incompatible with the \ScII \, line but still, even slight changes in
the adopted {\em rms} velocity will have an impact on the result.

\begin{figure}
  \centering
  \includegraphics[width=0.5\textwidth]{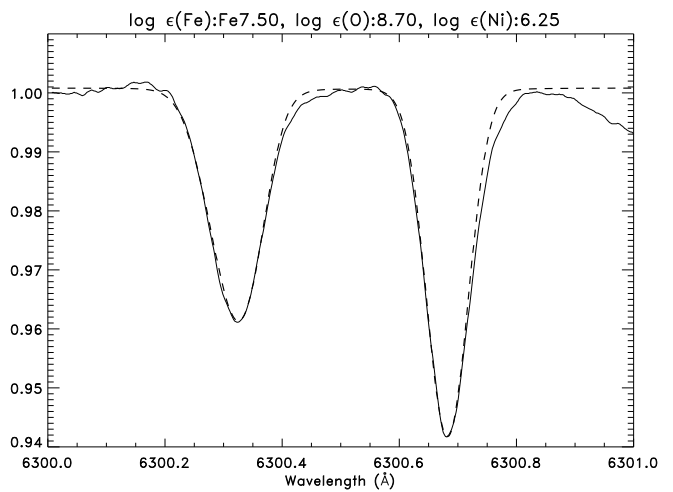}
  \caption{Fits with a low O abundance obtained after minor tweaks to
    the model atmosphere and the continuum reference selection. The
    goodness of this fit illustrates the high sensitivity of abundance determinations to the particular details of the model or the procedure.}
  \label{fig:lowabfit}%
\end{figure}

Playing with all of the above factors, it is even possible to achieve
an O abundance compatible with that of AGSAK. The actual fit is shown
in Fig~\ref{fig:lowabfit}. Notice how the fit is still remarkably
good, given how discrepant these new abundances are compared to the
previous figures, which again illustrates how extremely sensitive
these results are to small details of the model or of the procedure in
general. In order to produce this low abundance, the velocity
enhancement at the base of the photosphere was reduced by a 15\%
(bringing the {\em rms} velocity at that layer down from 1.36 to 1.15
km~s$^{-1}$), the wavelength scale was shifted 2~m\AA \ and the
continuum reference was decreased by 10$^{-3}$. The \ScII \ line
became slightly narrower and deeper with the change in the model,
creating a very noticeable discrepancy at the line core. This was
compensated by decreasing its abundance/$\log(gf)$ by 6\%. In this
manner, the fit to the \ScII \ line is slightly worse than it was
before but, as Fig~\ref{fig:lowabfit} shows, it might still be
considered acceptable.

\section{Discussion}

This paper presents a novel approach to the O abundance determination
using a 3D model that has been obtained from observations, as opposed
to most recent works that use a hydrodynamical simulation. Since the
model is obtained by inverting the much stronger \FeI \ lines at
6301.5 and 6302.5 \AA \ observed by the Hinode SP, it seems to be
missing some of the dynamics that take place in the deeper formation
region of the O/Ni blend. This could be due to a lack of spatial
resolution or to the weak sensitivity of the stronger \FeI \ lines to
plasma motions in a layer very close to the continuum formation
region. \cite{A08} already gave a hint that this could be a problem
and noted, with remarkable perspicacity, that ``a possible
disadvantage is that the derived motions will depend on the
instrumental resolution and might not be as vigorous as in fully
resolved 3D theoretical models''. Fortunately, this issue can be
overcome by using the nearby \ScII \ line, which is of very similar
strength to the O/Ni blend and probes the same atmospheric region.

In principle, the results obtained are in the high range
(\leO=8.9$\pm$0.1), compatible with the ``old'' solar
composition. However, in view of how sensitive the results are to the
details of the model atmosphere (it is even possible to produce values
compatible with the ``new'' composition with very minor tweaks), I
believe they should be taken with some reservation. In fact, the same
caution should probably be adopted with all previous results on
abundance determinations based on the traditional technique of fitting
atlas observations with a prescribed model atmosphere. Ultimately, the
reliability of our abundances is limited by the uncertainty in the
models employed, of which we still lack such detailed
understanding. It is then almost impossible to ascribe meaningful
error bars to those measurements. A thorough study of the systematic
errors arising from the use of a prescribed model and other parameters
employed in the abundance analysis is urgently needed in order to
gain a better understanding of our limitations.

An important conclusion to draw from this study is that one does not
necessarily obtain the ``low'' O abundance of AGSAK (\leO=8.66) when
using a 3D model with recent atomic data. Intermediate
(\citealt{CLS+08}) and high (this work) abundances may be obtained
from 3D models, as well. Furthermore, given the sensitivity of the
results to the details of the analysis, we should also question to
what degree an analysis based on equivalent widths (which is what has
been employed in most cases thus far) is enough to capture all the
subtleties that are observed in a detailed fit. It seems that in
difficult situations, such as this one, it is necessary to work with a
full line analysis quantified by some merit function (e.g., the
$\chi^2$ used here).

In the solar case there is an obvious way to improve the overall
procedure and increase our confidence in the measurement of element
abundances. Using spatially-resolved observations, it is pososible to
condcut a pixel-by-pixel analysis, similarly to how it was done in
\cite{SNN07}. If one has systematic errors, there is no reason to
expect that all of the analyzed spectra will yield consistent
results. Therefore, the pixel-to-pixel distribution is a perfect
sanity check to diagnose possible systematic errors. In that paper we
found a fluctuating pattern of abundances, correlated with the
granulation distribution. This inconsistency was probably a reflection
of the uncertainties in the non-LTE line formation physics. Conducting a
similar study with simpler lines such as the forbidden transition
analyzed here would undoubtly cast some more light into this important
issue.

\begin{table*}
  \caption[]{Spectral line data, where $r_0$ is the B\"ohr radius,
    $\gamma_{rad}$, $\gamma_{Stark}$ and $\gamma_{vdW}$ are the
    radiative, Stark and van der Waals damping enhancement constants.}
  \label{table:atomic}
  $$ 
  \begin{array}{ccccccccc}
    \hline
    \noalign{\smallskip}
  Element & \lambda \  (\mbox{\AA})  & Excitation (eV)  & log(gf) & \sigma (r_0^2) 
  &  \alpha & \gamma_{rad} & \gamma_{Stark} & \gamma_{vdW} \\
    \noalign{\smallskip}
    \hline
    \noalign{\smallskip}
  \OI  &  6300.304 & 0.000 & -9.819  & - & - & 0.0 & 0.05 & 1.00 \\
  \NiI &  6300.335 & 4.266 & -2.253  & - & - & 2.63 & 0.054 & 1.82 \\
  \NiI &  6300.355 & 4.266 & -2.663  & - & - & 2.63 & 0.054 & 1.82 \\
  \ScII & 6300.678 & 1.507 &  -1.89  & - & - & 2.30 & 0.050 & 1.30 \\
  \FeI &  6301.501 & 3.654 & -0.718  & 834.4 & 0.243 & - & - & -  \\
  \FeI &  6302.494 & 3.686 & -1.160  & 850.2 & 0.239 & - & - & -  \\
    \noalign{\smallskip}
    \hline
  \end{array}
  $$ 
\end{table*}

\begin{table*}
  \caption[]{Inversion nodes}
  \label{table:nodes}
  $$ 
  \begin{array}{lccccccccc}
    \hline
    \noalign{\smallskip}
     & \vline &         & Magnetic \, pixel &        &    &     & \vline
     &  Non-magnetic \, pixel &  \\
   Physical  & \vline &  Comp 1 &   & \vline &Comp 2  &    & \vline &        \\
   Parameter  & \vline & Cycle 1 & Cycle 2 & \vline& Cycle 1 & Cycle 2 & \vline & Cycle 1 & Cycle 2 \\
    \noalign{\smallskip}
    \hline
    \noalign{\smallskip}
  Temperature &        &  3  &  5  & &     2   &   3  & &  3 &  5 \\  
  L.o.s. \,  velocity    &        &  1  &  3  & &     1   &   2  & &  2 &  3 \\     
  Microturbulence &    &  1  &  1  & &     0   &   0  & &  0 &  0 \\
  B_{long}     &        &  0  &  0  & &     1   &   2  & &  0 &  0 \\
  B_x         &        &  0  &  0  & &     1   &   2  & &  0 &  0 \\
  B_y         &        &  0  &  0  & &     1   &   2  & &  0 &  0 \\
  Filling \, factor &     &  0  &  0  & &     1   &   1  & &  0 &  0 \\
    \noalign{\smallskip}
    \hline
  \end{array}
  $$ 
\end{table*}
\begin{acknowledgements}

Hinode is a Japanese mission developed and launched by ISAS/JAXA,
collaborating with NAOJ as a domestic partner, NASA and STFC (UK) as
international partners. Scientific operation of the Hinode mission is
conducted by the Hinode science team organized at ISAS/JAXA. This team
mainly consists of scientists from institutes in the partner
countries. Support for the post-launch operation is provided by JAXA
and NAOJ (Japan), STFC (U.K.), NASA, ESA, and NSC (Norway). 

The author thankfully acknowledges the technical expertise and
assistance provided by the Spanish Supercomputing Network (Red
Española de Supercomputación), as well as the computer resources used:
the LaPalma Supercomputer, located at the Instituto de Astrofísica de
Canarias.

Financial support by the Spanish Ministry of Science and Innovation
through project AYA2010-18029 (Solar Magnetism and Astrophysical
Spectropolarimetry) is gratefully acknowledged.

The VALD database (referenced in the text) was used to obtain atomic
parameters for the \ScII \ line.

\end{acknowledgements}




\end{document}